# A Survey on Quantitative Modeling of Trust in Online Social Networks

WENTING SONG, University of Texas at Austin, USA

K. SUZANNE BARBER, University of Texas at Austin, USA

Online social networks facilitate user engagement and information sharing but are also rife with misinformation and deception. Research on trust modeling in online social networks focuses on developing computational models or algorithms to measure trust relationships, assess the reliability of shared content, and detect spam or malicious activities. However, most existing review papers either briefly mention the concept of trust or focus on a single category of trust models. In this paper, we offer a comprehensive categorization and review of state-of-the-art trust models developed for online social networks. First, we explore theories and models related to trust in psychology and identify several factors that influence the formation and evolution of online trust. Next, state-of-the-art trust models are categorized based on their algorithmic foundations. For each category, the modeling mechanisms are investigated, and their unique contributions to quantitative trust modeling are highlighted. Subsequently, we provide an implementation-centric trust modeling handbook, which summarizes available datasets, trust-related features, promising modeling techniques, and feasible application scenarios. Finally, the findings of the literature review are summarized, and unresolved challenges are discussed.

## 1 INTRODUCTION

Online Social Networks (OSNs) are digital platforms that facilitate the creation, sharing, and exchange of information, ideas, and content among users. Nowadays, people tend to rely on online social networks and online community applications for daily social interactions. OSNs simplify the process for people to participate in online social activities and eliminate restrictions regarding when and where content can be created and shared. Nevertheless, online social networks are also rife with disinformation, misinformation, and deception [187]. User-generated content in online social networks exhibits significant variation in quality due to a variety of factors. Online users come from diverse backgrounds and have varying levels of expertise in the topics they discuss. Some may provide accurate and insightful information, while others may lack expertise and unknowingly share misleading information. Additionally, users often have personal biases and subjective opinions,

Authors' Contact Information: Wenting Song, University of Texas at Austin, Austin, Texas, USA, wentingsong@utexas.edu; K. Suzanne Barber, University of Texas at Austin, Austin, Texas, USA, sbarber@identity.utexas.edu.

which may affect the quality of their contributions. The motivation behind creating content also affects its quality. Some users may intentionally create low-quality or misleading content for reasons such as trolling, spreading propaganda, or creating clickbait. In recent years, the emergence of generative AI models such as Generative Adversarial Networks (GANs) [48] or Variational Autoencoders (VAEs) [78] has led to a surge in online content. With the massive amount and diversity of user-generated and AI-generated content, it is becoming increasingly difficult, if not impossible, for users to distinguish authenticity and find valuable information. With all these issues in mind, it is imperative to critically evaluate online content by examining its source, accuracy, and relevance.

Trust is the bedrock of any relationship or connection, whether online or offline. Online trust plays a vital role in developing meaningful connections in online social networks. When trust is established, users are more likely to connect with others and cultivate reliable and productive interactions. The rise of trustworthy AI research emphasizes several trust-related factors, including robustness, generalization, explainability, transparency, repeatability, fairness, accountability, privacy, and security [77, 88]. Current research on trustworthy AI primarily focuses on establishing guidelines and standards for AI technologies [99]. In contrast, trust modeling in online social networks focuses on developing computational models and algorithms to evaluate trust relationships, verify the credibility of information, detect malicious or spam activities, and cultivate more trustworthy and high-quality online communities.

Research on trust modeling in online social networks is booming, but a systematic review of state-of-the-art trust models is still lacking. We are working to address this gap. In this survey paper, we will review the latest trust models developed for online social networks and explore the following questions:

- **RQ1**: What is trust? How is trust defined in online social networks?
- **RQ2**: What information or data is available to compute online trustworthiness?
- **RQ3**: How is online trust quantified?
- **RQ4**: What application scenarios benefit from trust quantification models?

Figure 1 provides an overview of this survey paper. First, we unpack the concept of trust and examine relevant psychological theories and models. Drawing on these theories, we identify key factors that influence the formation and evolution of trust, laying the foundation for trust modeling. Next, we examine trust models from different application domains to gain a more comprehensive understanding of the role of trust in online social networks. The state-of-the-art trust models are categorized into ten representative algorithmic categories. Throughout the literature review, we emphasize the practical implementation of trust models, focusing on data prerequisites, algorithmic foundations, and contextualized application scenarios. Thereafter, we provide an implementation-centric trust modeling handbook to inspire the development and implementation of trust models.

In summary, the **main contributions** of this paper are as follow:

- The concept of trust is unpacked, and key factors influencing the formation and evolution of online trust are identified.
- A comprehensive categorization and review of state-of-the-art research on trust modeling in online social networks.
- An implementation-centric trust modeling handbook summarizing available datasets, trust-related features, promising modeling techniques, and feasible application scenarios.

The paper is organized as follows. Section 2 describes related work in terms of other surveys reviewing trust modeling. The definition and influencing factors of online trust are explored in Section 3. In Section 4, we review the development and evolution of trust models over the past two decades and discuss their trends. Section 5 provides a literature review of several representative state-of-the-art trust models developed for online social networks. The summary table in Section 6 outlines the data prerequisites, algorithmic foundations, and application scenarios of these trust models. In Section 7, we provide an implementation-centric trust modeling handbook, which summarizes available datasets, trust-related features, promising modeling techniques, and feasible application scenarios. The findings are discussed in Section 8. Lastly, Section 9 highlights the open challenges and research directions and Section 10 concludes the paper.

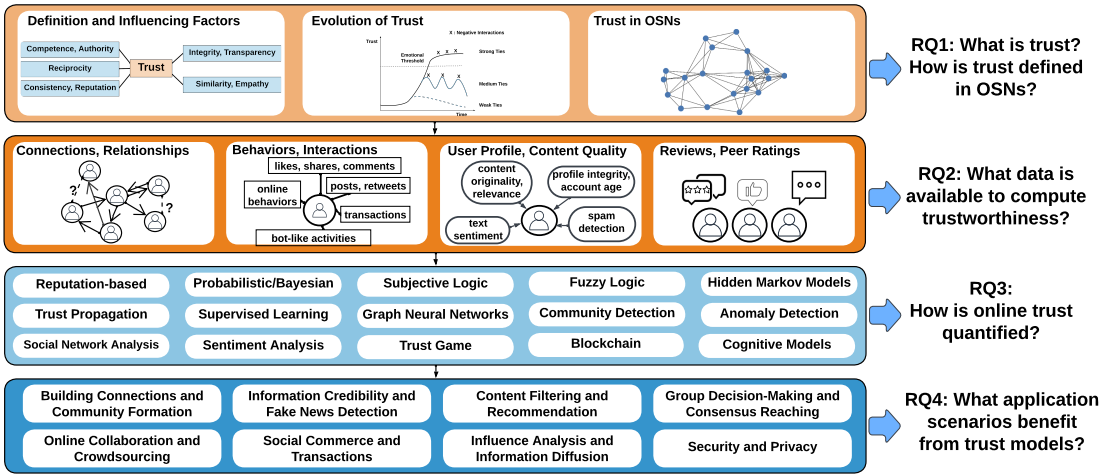

Fig. 1. The overall structure of this survey paper.

## 2 RELATED WORK

There are literature reviews on trust modeling in various fields, including agent systems [132], cybersecurity [56], the Internet of Things [5, 41], and online behavior research [54, 186]. Krausman et al. [81] reviewed trust measurement principles and proposed a methodological toolkit to facilitate the development of new trust models for human-autonomy teams [64]. Fan et al. [38] studied trust aggregation models and analyzed the resilience of trust metrics in the face of common threats in decentralized trust management systems. This review focuses on trust modeling in online social networks. First, we unpack the concept of trust and explore theories and models related to trust in psychology. Drawing on these theories, we identify several factors that influence the formation and evolution of online trust, laying the foundation for trust modeling. In this review, we focus on computational models that shed light on how trust is formed and evolves in online communities.

According to our study, a comprehensive literature review of the latest trust models in online social networks is still lacking. Some were published several years ago and may not capture the latest advances [23, 139]. Some delve into specific categories of trust models [16, 44, 69, 156], lacking a holistic perspective. This survey paper aims to fill this gap. We conduct a systematic review of the state-of-the-art trust models developed for online social networks and categorize these models into ten categories based on their algorithmic foundations. Furthermore, we observe that existing trust survey papers tend to propose the concept of trust but fail to elaborate on how to implement a trust model or how to apply the trust concept in practice [186]. In response, this review places particular emphasis on the development and implementation of trust models. In addition, we provide an implementation-centric trust modeling handbook that summarizes available datasets, trust-related features, promising modeling techniques, and feasible application scenarios.

# 3  DEFINITION AND FACTORS OF TRUST

This section focuses on the question: What is trust? How is trust defined in online social networks? Given that trust is essentially a psychological state, we first examine relevant theories and models in psychology. These theories offer valuable insights when modeling the establishment and development of trust within online communities. Next, we decompose the concept of trust and identify several factors that influence the evolution of online trust, laying the foundation for trust modeling.

## 3.1  What is Trust?

Julian Rotter defined trust as a general expectation that the words, promises, or statements of others can be relied upon [131]. Mayer et al. [110] define trustworthiness as a function of competence, benevolence, and integrity. As a psychological state, trust is a willingness to accept vulnerability based on positive expectations of another's behavior [110]. Moreover, Blau's social exchange theory [8] argues that trust develops through repeated social interactions with the expectation of mutual benefit. Lewicki and Bunker [87] proposed three stages of trust evolution over time: calculus-based trust, knowledge-based trust, and identification-based trust. Trust often begins with calculus-based trust, where individuals weigh potential risks against expected benefits. Through repeated interactions and positive experiences, trust evolves into knowledge-based trust. Ultimately, emotional connections and shared values foster identification-based trust, which can be seen as deeper relationships and stronger community bonds [111].

## 3.2  Trust Influencing Factors

Drawing on psychological theories and models related to trust, we highlight several factors that influence the formation and evolution of trust. As shown in Figure 2, the diversity of trust factors also highlights the multifaceted nature of trust.

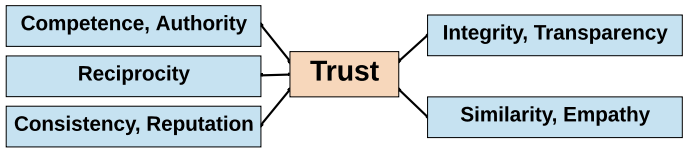

Fig. 2.  Trust Influencing Factors

- Competence, Authority
  Competence is emphasized as a key factor in various theories of trust [30, 110, 112]. Trust is established when a trustee demonstrates a high level of competence in achieving desired outcomes or is recognized as an authority in a field. If trustors perceive the trustee as knowledgeable or efficient, they will quickly develop trust [112].

- Reciprocity
  According to Deutsch's theory [30], trust is a decision made under uncertain conditions with potential risks and benefits. Trustors tend to assess the risks involved in an interaction and determine whether the potential benefits outweigh the risks [107, 186]. Trust grows when both parties are fair and reciprocal (Blau's social exchange theory [8]).

- Consistency, Reputation
  Trustees are expected to demonstrate consistent, reliable behavior over time [131]. In online social networks, users' reputation is usually determined by their past behavior and numerical indicators such as reviews, ratings, or peer endorsements. Predictable behavior enhances the perception of trustworthiness.

- Integrity, Transparency
  Integrity, transparency, and clear communication are essential to building lasting trust. [110]. Trust is built when the other party is open about their intentions, motivations, and information. For example, health apps are more likely to earn users' trust if they openly inform users about how their data is used and demonstrate a commitment to transparency and ethical practices.

- Similarity, Empathy
  Trust stems from shared experiences, emotional connections, and empathy [111]. This emotional bond and shared values foster identification-based trust [87], which builds deep and lasting relationships.

### 3.3 Characteristics of Online Trust

In face-to-face interactions, trust is built through shared experiences and social cues such as eye contact and body language. In the absence of physical cues, online trust is primarily established through digital cues, including but not limited to reputation indicators (e.g., reviews, ratings, and user endorsements) and profile indicators (e.g., photos, bios, affiliations, and verified credentials). This section discusses the characteristics of online trust.

(1) Context Dependency and Domain Specificity
Trust varies by platform, community norms, and type of interaction [49]. A user might trust someone on LinkedIn for professional advice, but not someone on Instagram for financial advice.

(2) Trust Maintenance or Decay
Trust is built gradually through repeated positive interactions [87, 131]. Meanwhile, trust can decay quickly due to negative interactions or violations (e.g., deception, manipulation) [142] (Figure 3).

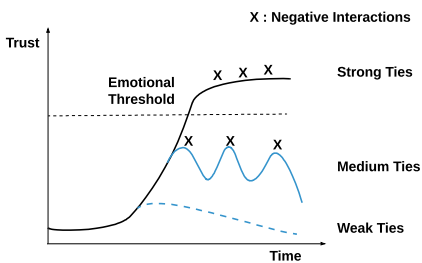

Fig. 3. **Development and Decay of Trust Over Time** [142]. Three types of ties are illustrated. Strong ties mean that trust is built through multiple interactions in a relatively short period of time, breaks through emotional thresholds, and can withstand negative interactions. Moderate ties mean that trust develops through fewer interactions, has not yet reached an emotional threshold, and will weaken if interactions cease or negative interactions persist. In weak ties, negative interactions or lack of interaction can lead to the end of the relationship.

(3) Anonymity and Iduser Cues
Online users often participate in a partially or fully anonymous manner, which makes building trust more difficult. While real names can enhance trust, anonymity may encourage more open and honest expression. Additionally, group affiliations or verification badges can serve as supplementary indicators of trustworthiness.

(4) Trust Propagation
   Trust is transitive and can be transferred across connections, albeit with some reduction. Trust propagation is a fundamental component of e-commerce systems (e.g., eBay), aiming to help users find trustworthy reviewers, sellers, and products.

(5) Cognition and Affectiveness
   McAllister [111] argues that trust includes both rational evaluation (cognition-based trust) and emotional bonds (affect-based trust). Cognitive trust stems from rational judgment and observable evidence, such as evaluating someone's competence or contributions. Affective trust, on the other hand, stems from emotional connections, including empathy, intimacy, and shared values [87].

(6) Interpersonal Trust and System-based Trust
   Interpersonal trust refers to trust in a specific individual. In contrast, system-based trust [47], also known as institutional trust, refers to the trust that individuals place in institutions (e.g., online platforms). It encompasses users' trust in the platform's ability to manage moderation, protect privacy, and regulate content (e.g., flagging or removing harmful content).

## 4   EVOLUTION OF TRUST MODELS

Before delving into specific trust models, we will first review the evolution of trust models over the past two decades. Figure 4 provides an overview of the evolution of trust models developed for online social networks since the 2000s.

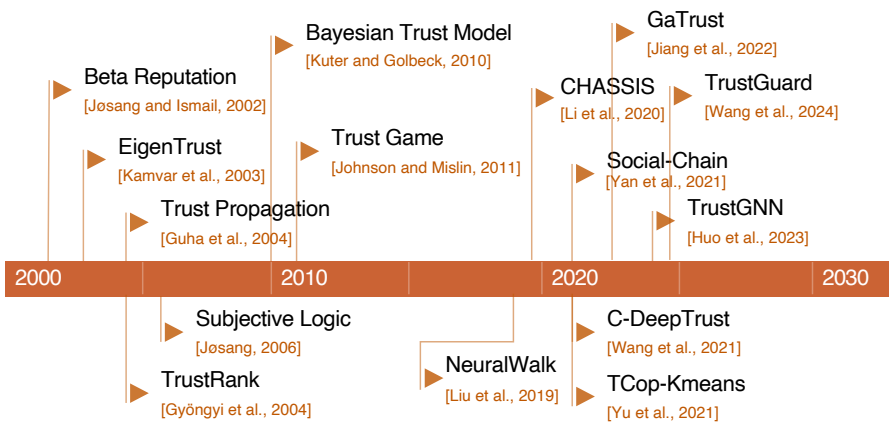

Fig. 4. The evolution of trust models developed for online social networks from the 2000s to the 2020s.

Initially, trust modeling was an interdisciplinary study that combined psychology and human factors [110, 131]. In the 2000s, research on trust modeling in online social networks drew inspiration from the conceptual definition of trust and focused on modeling the basic attributes of trust. Probabilistic and subjective logic trust models have been proposed to model the subjectivity and uncertainty in trust relationships [22, 72, 73]. In reputation trust models [55, 75], users accumulate trust scores through their behavior, contributions, and peer feedback during interactions. Various trust models have collectively enhanced our understanding of how online trust is established, maintained, or destroyed.

Over the past decade, the field has gradually shifted towards data-driven and machine learning approaches. With the development of deep learning models and the continuous emergence of large-scale datasets, many trust models have adopted techniques such as feature-based supervised learning models [156], Graph neural networks [68, 93], K-means clustering [176], and anomaly detection [163] to quantify trust. In recent years, the techniques and application scenarios of trust models have become increasingly diversified. Decentralized trust evaluation models based on blockchain [38, 150], cognitive trust models [90], trust models developed for misinformation detection [71, 141], and trust models for human-AI collaboration [148] have also emerged.

## 5 A SYSTEMATIC REVIEW OF TRUST QUANTIFICATION MODELS

This section systematically reviews the state-of-the-art research on trust modeling and categorizes the models into ten categories based on their algorithmic foundations. For each category, we investigate the underlying modeling mechanisms and highlight their unique contributions to the quantitative modeling of trust in online social networks.

### 5.1 Reputation-Based Trust Models

Early research on trust models focused on reputation-based models (e.g., EigenTrust [75] and TrustRank [55]). User trustworthiness is measured as a reputation score, which is accumulated through behavior, contributions, and peer feedback during interactions. Reputation-based trust models aggregate community opinions, and the computation process is transparent to users. However, such models are vulnerable to Sybil attacks [118], collusion, and fake reviews.

Rezvani and Rezvani [128] proposed a novel reputation system to distinguish fake and trustworthy consumer ratings. Inspired by the iterative filtering algorithm [29], users whose ratings frequently deviate from those of other users will be assigned lower weights. The proposed random iterative filtering algorithm simultaneously calculates user weights and aggregates reputation scores. Using only a random subset of ratings, it can approximate the average rating for each product within a certain error threshold. This randomness also makes the system resistant to unfair ratings and simple attacks.

GOLI [65] aims to identify opinion leaders in online social networks by evaluating user reputation. Initially, it constructs a trust network by analyzing user characteristics and social connections. Subsequently, the reputation score is derived from feedback during interactions and trust ratings provided by neighbors. Finally, the aggregate reputation score is combined with in-degree centrality to identify influential opinion leaders.

Wu et al. [164] developed an optimal feedback model to prevent manipulation during group decision-making. Based on trust connections, each expert is assigned a behavioral weight, which determines their influence in reaching consensus. In addition, the model calculates optimal opinion adjustment costs to promote consistent behavior.

### 5.2 Probabilistic and Bayesian Trust Models

Probabilistic trust models represent trust as a probability distribution rather than a fixed number [83]. They handle the uncertainty of trust and have a solid mathematical foundation. For instance, Bayesian trust models use Bayesian inference to update the probability of a user being trustworthy based on new evidence or interactions [14, 25, 161]. Probabilistic trust models work well with sparse interactions or noisy data. However, they require prior distributions and can be computationally expensive.

Wang et al. [160] are committed to curbing the spread of negative information through intervention strategies. They defined four user states and calculated the state transition probabilities based on trust relationships. Three collaborative intervention strategies—warning, correction, and

guidance—were introduced to address users in different states: the unknown state, the negative information diffusion state, and the dual information hesitation state. Moreover, Naumzik and Feuerriegel [117] employed stochastic processes [43, 127] and Bayesian inference to model information diffusion and detect misinformation on social media. Inspired by the Hawkes process [57, 185], they developed a new probabilistic mixture model that can simulate retweet cascades and directly classify them as true or false, eliminating the need for feature engineering.

Lingam et al. [94] worked on social bots and botnet community detection on Twitter (X). First, a weighted graph was constructed, where nodes represent accounts and edges represent behavioral similarities and trust scores between accounts. Trust scores were calculated using Bayesian inference [25] and random walk stochastic processes [66]. Next, the authors leveraged graph theory and deep autoencoders to identify communities (groups of accounts) engaging in malicious behavior. Research also shows that bots tend to grow followers more slowly than real accounts and have a disproportionate follower-to-following ratio.

### 5.3 Subjectivity and Uncertainty-Based Trust Models

Subjective logic is a probabilistic logic that deals with the uncertainty in trust relationships [22, 72, 73]. Trust is represented by an opinion triple consisting of belief (b), disbelief (d), and uncertainty (u). Fuzzy logic deals with the ambiguity in trust using linguistic terms (e.g., "high trust" or "low trust") and membership functions [177, 178]. They are useful in situations where trust is not easily quantifiable. However, the definitions of opinion triples and membership functions may vary in different contexts and contain biases [58, 121].

Cho et al. [24] aims to reduce the spread of false information. They proposed an enhanced opinion model based on subjective logic to classify user opinions on false information into trust, distrust, and uncertainty. The simulation experiments revealed several factors that influence the spread of false information, including the ratio of information propagation frequency, centrality measures, opinion decay rate, and the proportion of true informants.

Ma et al. [105] proposed a novel trust-based consensus-reaching model for group decision-making. Intuitive fuzzy sets deal with uncertainty and ambiguity in the interaction between experts. The model assigned weights to experts using a second-order additive fuzzy measure combined with the Shapley value. Experts with higher weights had more impact on reaching a consensus. One limitation is that the model assumes that social trust relations are static.

3VSL-AT [97, 98] was designed for trust computation in online social networks. The authors proposed a three-valued subjective logic model to capture uncertainty in trust and used Bayesian inference to calculate trust relationships. Leveraging depth-first search (DFS), the AssessTrust algorithm can calculate the trust relationship between any pair of users.

### 5.4 Context-Aware Trust Models

Trust is situational, not universal. A user might trust someone for movie recommendations but not for financial advice. Context-aware trust models calculate trust scores specific to domains, topics, or situations rather than taking a one-size-fits-all approach [158, 187]. It is useful for personalized recommendations, content curation, and contextual decision support. The challenges come from the need to detect multiple contexts, increased data requirements, and more complex modeling.

DDTrust [45] was proposed as a dynamic trust relationship prediction model between user pairs. The authors employed two text analysis tools (LIWC [15, 124] and BERT [32]) to identify user sentiment and generate word embeddings from tweet contexts and user profiles. It improves trust relationship prediction by combining user sentiment analysis and textual content inspection. Similarly, DCAT [46] predicts pairwise trust relationships based on user demographics and textual content. LIWC [15, 124] and word embedding techniques were used for contextual analysis. In addition, Gatrust [68] integrates node attributes, contextual features, network topology, and trust interactions into a graph neural network approach to improve the accuracy of trust relationship prediction.

Xue et al. [171] proposed a novel spam review detection model based on trust evaluation. The model evaluated user trustworthiness by analyzing review content and examining the opinions expressed. Key evaluation factors include the inconsistency of opinions in user reviews and the extent to which these reviews deviate from the majority. The study found that user trustworthiness can be inferred from the opinions they express in online reviews.

## 5.5 Game-Theoretic Trust Models

The core idea of game-theoretic trust models is that trust emerges from rational choices in repeated interactions. They model trust as the outcome of strategic interactions and apply game theory principles to analyze the interactions between users [59, 70, 143]. Popular game-theoretic trust models include the prisoner's dilemma [82, 125, 126], Bayesian games, and signaling games [21]. It is well-suited for market-oriented social networks where incentives matter. Nonetheless, many platforms allow users to remain anonymous or create multiple idusers, reducing the effectiveness of repeated gaming strategies or penalties for breach of trust.

Mahajan et al. [106] proposed a novel trust-based team formation algorithm named PRADA-TF. It modeled the team formation process as a strategic game and adopted the Mechanism Design algorithm [12, 119]. PRADA-TF selected team members by evaluating their trustworthiness. The team formation process prioritizes domain expertise and user privacy preferences, aiming to optimize team performance while respecting individual privacy preferences.

To combat the spread of disinformation, Guo et al. [52, 53] introduced an opinion framework using repeated games. They modeled the interaction as a three-player game involving attackers, users, and defenders. The authors exploited subjective logic to model how users update their opinions under uncertainty and how social interactions influence the spread of disinformation. Similarly, Zhang et al. [180] proposed an evolutionary information diffusion model based on Evolutionary Game Theory (EGT) [4, 168], aiming to mitigate the adverse effects of malicious users.

## 5.6 Graph-Based Trust Models

Graph-based trust models describe social networks as weighted graphs, where users are represented as nodes and trust relationships are represented as weighted edges [34, 69, 108]. Trust is propagated or aggregated across network paths [51]. These models exploit network structure and trust transitivity to infer trust from connections and paths [170]. Typical use cases include friend recommendations and spammer detection. Graph-based trust models can effectively capture structural relationships, but require known explicit trust relationships to function.

The most common application of graph-based trust models is to predict trust relationships [93, 96, 159, 169]. Guardian [93] was introduced based on graph neural networks (GNNs), which learns the latent factors of trust propagation and trust aggregation. It predicts the trust relationship between any two users by examining known trust relationships and social network structures. To improve the accuracy of trust relationship prediction, AtNE-Trust [159] captures the characteristics of trust networks and multi-dimensional user attributes simultaneously. With the same goal, AHNTP [17] adopts hypergraphs to model high-order dependencies and integrates motif-based PageRank [184] to identify social influencers and enhance trust propagation.

Wu et al. [165] leveraged trust propagation models to calculate indirect trust relationships and construct a complete trust network. The calculated trust score serves as a weighting factor, simplifying and reducing the effort required to reach consensus in group decision-making.

### 5.7 Machine Learning and Data-Driven Trust Models

Machine learning trust models use features of interactions and network structure to learn and predict trustworthiness [156]. Supervised learning models rely on labeled trusted/untrusted data for training. Unsupervised learning models such as clustering and anomaly detection can identify unusual behavior patterns that may indicate untrustworthiness [163]. Graph neural networks (GNNs) learn trust embeddings directly from the social graph for trust prediction [68, 93]. Meanwhile, machine learning trust models are prone to bias, and some lack interpretability due to their complex structures.

PTP-MF [113] adopts matrix factorization to predict pairwise trust relationships. Each user is represented by a feature vector pair $\mathbf{P}_i$ and $\mathbf{Q}_i$. The trustor vector $\mathbf{P}_i$ represents the user's tendency to trust others. The trustee vector $\mathbf{Q}_i$ represents how others perceive the trustworthiness of the user. Experiments showed that PTP-MF can make accurate predictions even when the user connection graph is sparse.

Graph neural networks (GNNs) have emerged as a powerful approach for learning complex relationships and dependencies in graph-structured data [135, 153, 166, 167]. TrustGNN [63] is a representative GNN-based trust model that achieves satisfactory trust prediction accuracy and exhibits robustness against malicious users. The authors combined the graph attention mechanism [153] with trust propagation theory and proposed a trust-specific message passing where the attention weights reflect the trustworthiness. Trustguard [157] is another trust model that utilizes GNNs and a position-aware attention mechanism [183]. Differently, it captures trust dynamics and integrates defense mechanisms within the spatial aggregation layer, enhancing its ability to resist trust-related attacks.

TCop-Kmeans [176] is a Trust Cop-Kmeans clustering algorithm designed to support consensus reaching in large-scale decision making. It leveraged the similarity in opinions and trust relationships among decision-makers to optimize the clustering algorithm. Additionally, the trust relationship scores were used as weights to develop a minimum cost consensus model.

### 5.8 Blockchain and Decentralized Trust Models

Blockchain enables decentralized trust computation, immutability of interactions, and transparency in trust relationships [38, 56, 67, 150]. All transactions or interactions are recorded on an immutable, timestamped distributed ledger. Trust evaluation is conducted in a distributed manner on the blockchain, making the system more robust and reducing the risk of malicious tampering or Sybil attacks [133]. Based on blockchain technology, Sarode et al. [133] proposed a novel decentralized social network. The immutability of blockchain ensures that all transactions and content remain tamper-proof, protecting them from unauthorized modifications. In addition, the system incorporates incentive mechanisms to promote honest and reliable participation—users can be rewarded for creating valuable and accurate content, while malicious behavior will be punished.

Social-Chain [173] was proposed as a new blockchain-based decentralized trust evaluation model. The authors introduced a new consensus mechanism, Proof of Trust (PoT), where miners can generate new blocks by collecting new trust evidence. This trust evaluation system employed a distributed architecture, making it resistant to traditional attacks.

Yu et al. [175] proposed a distributed reputation mechanism to distinguish the authenticity of information in online social networks and prevent the spread of misinformation. User reputation was assessed based on the authenticity of the messages they sent. The authors used a Transformer-based language model (RoBERTa [102]) combined with sentiment analysis techniques to verify the authenticity of the messages. Blockchain smart contracts were utilized to automatically calculate and update user reputation scores.

### 5.9 Psychological and Cognitive Trust Models

McAllister [111] believes that building trust includes both rational evaluation (cognitive trust) and emotional bonds (affective trust). Cognitive trust models draw on psychology and cognitive science to simulate how users perceive, form, and update trust when interacting with other users or content in online social networks [19, 37, 39]. Granatyr et al. [50] argued that considering cognitive factors is crucial when constructing a trust model.

CHASSIS [90] applied the Hawkes process [57, 185] to simulate information diffusion. The authors incorporated conformity theory from social psychology into the information diffusion model [3, 26, 31] and quantified conformity by examining diffusion trees [89]. The introduced semi-parametric inference method helps identify which individuals exhibit conformity behavior.

Karami et al. [76] studied the characteristics and motivations of users who spread misinformation on social media. Drawing on psychological theories [33, 147], they identified five motivational factors: uncertainty, anxiety, lack of control, relationship enhancement, and social rank. The motivational factors help distinguish misinformation spreaders from ordinary users and enable the early detection and crackdown on the spread of misinformation.

### 5.10 Hybrid and Multi-Component Trust Models

The multi-component trust model assesses trust by comprehensively considering multiple factors. It leverages the advantages of multiple methods to achieve a more nuanced, robust, and accurate trust evaluation.

TrustLab™ [61, 62, 179] was introduced to measure individual trustworthiness on social media. Trust was assessed based on six trust attributes: authority, experience, expertise, iduser, proximity, and reputation. It calculated user trust scores from multiple dimensions, including user characteristics, historical posting behavior, domain expertise, physical proximity, and social network structure. Additionally, various charts were used to visualize the distribution of Twitter (X) user trust scores.

T-VLC [20] was proposed as a general trust model for assessing trust among participants in online learning communities. It measures trust across eight dimensions: direct experience, reputation, role, knowledge, security, quality, institutional trust, and closeness, with each dimension assigned an aggregated weight. As a flexible framework, the model allows these weights to be adjusted to suit different application scenarios.

iSim [42] was proposed as a multi-component trust prediction method based on user similarity. This method employed techniques such as vector space similarity, trust propagation, and time-aware matrix factorization for comprehensive trust assessment.

## 6 REPRESENTATIVE TRUST MODEL REVIEW

In this section, we summarize several state-of-the-art trust models (TM) for online social networks introduced between 2019 and 2025 in Table 1. Throughout the literature review, we emphasize the implementation aspects of trust modeling, focusing on data prerequisites, algorithmic foundations, and contextualized application scenarios.

Table 1. A Review of the State-of-the-Art Trust Models (TM) for Online Social Networks (2019-2025).

| | Models | Data Requirements | Algorithms/Techniques | Application Scenarios |
|---|---|---|---|---|
| **1. Reputation-Based TM** | Rezvani and Rezvani [128] (2020) | user-provided ratings | iterative filtering [29] | to distinguish between trustworthy and fake consumer ratings |
| | GOLI [65] (2023) | user characteristics, number of followers, user connections, posts, response time | graph neural networks, in-degree centrality | to classify node importance and identify opinion leaders in online social networks |
| | Wu et al. [164] (2020) | existing trust relationships between users, user online behaviors | social network analysis, in-degree centrality | to prevent individual and group manipulation during the consensus-reaching process of group decision-making |
| **2. Probabilistic/Bayesian TM** | Wang et al. [160] (2020) | existing trust relationships between users | state transitions, Pontryagin's maximum principle | to suppress the spread of negative information through collaborative intervention strategies |
| | Naumzik and Feuerriegel [117] (2022) | retweet cascades, user tweets, number of followers | Hawkes process [57, 185], Bayesian inference [43, 127], Markov chain Monte Carlo sampling [43] | to detect false information on social media; classify information as true or false based on the propagation process |
| | DA-SBCD [94] (2020) | user social activities (tweets, retweets, posts), user profile, number of followers | Bayesian theorem [25], random walk stochastic process [66], deep autoencoders, similarity measures, sentiment analysis | to detect botnet communities and groups of accounts with malicious behavior |
| **3. Subjectivity/Uncertainty TM** | Cho et al. [24] (2019) | user connections within online social networks, user online behaviors | subjective logic, centrality measures, opinion decay | to mitigate false information and fake news on social networks |
| | Ma et al. [105] (2021) | interactive relationships between experts | Choquet integral, intuitionistic fuzzy set, Shapley value, centrality measures | to facilitate the consensus-reaching process in group decision-making |
| | 3VSL-AT [97, 98] (2021) | user connections in online social networks | Bayesian inference, Dirichlet-categorical distribution, subjective logic (discounting and combining operators), depth-first search | to compute the trust relationship between any two users in an online social network |
| **4. Context-Aware TM** | Xue et al. [171] (2019) | user online reviews | SVM classifier, Word2Vec [115], three-layer trust propagation [154] | to detect spam reviews through trust evaluation; assess the trustworthiness of users, reviews, and statements by analyzing the opinions expressed in reviews |
| | DDTrust [45] (2020) | tweet context, user profile, number of followers | LIWC [15, 124], BERT [32], neural networks | to identify user emotions from tweet context; to predict trust relationships between user pairs over time |
| | Gatrust [68] (2022) | existing trust relationships between users, user contextual features | graph neural networks | to predict trust relationships between user pairs |

| | Models | Data Requirements | Algorithms/Techniques | Application Scenarios |
|---|---|---|---|---|
| **5. Game-Theoretic TM** | PRADA-TF [106] (2024) | user connections, user profile (domain expertise, privacy preferences) | Mechanism Design [12, 119], subjective logic (discounting and consensus operators), betweenness centrality | introduced a trust-oriented team formation algorithm that selects potential members by considering domain expertise and privacy preferences |
| | Guo et al. [52, 53] (2022) | user type (active user/fake bots/spammers), user interactions, tweets texts | Latent Dirichlet Allocation, subjective logic, Nash Equilibrium | proposed a game-theoretic opinion framework to study opinion dynamics and combat the spread of disinformation |
| | Zhang et al. [180] (2020) | user connections, user type (rational/malicious users) | evolutionary game theory, reputation mechanisms, social norm, death-birth updating rule | proposed an evolutionary model of information diffusion to mitigate the negative impact of malicious users |
| **6. Graph-Based TM** | Wu et al. [165] (2021) | existing trust relationships | social network analysis | construct a complete trust network using trust propagation models; calculated trust scores used as weights to promote consensus in group decision-making |
| | AHNTP [169] (2024) | existing trust relationships, user behaviors, user-provided ratings/reviews | hypergraphs [17], motif-based PageRank, cosine similarity [184] | to predict trust relationships; improve prediction accuracy by capturing high-order dependencies with hypergraphs |
| | AtNE-Trust [159] (2020) | existing trust relationships, user-provided ratings and reviews | skip-gram with negative sampling [115, 116], doc2vec [84], autoencoders, cosine similarity, matrix factorization | to predict pairwise trust relationships; combine trust network features and multi-dimensional user attributes to enhance trust relationship prediction |
| **7. Machine Learning TM** | PTP-MF [113] (2019) | existing trust relationships | matrix factorization, cognitive biases, gradient descent | to predict trust relationships between user pairs |
| | TrustGNN [63] (2023) | existing trust relationships | graph neural networks, attention mechanism [153] | to predict trust relationships; model various trust propagation patterns and construct different trust chains for a comprehensive trust evaluation |
| | TrustGuard [157] (2024) | temporal trust ratings from other users | graph neural networks, position-aware attention mechanism [183], trust propagation, similarity measures | to predict trust relationships; model trust dynamics; incorporate defense mechanisms to combat trust-related attacks |
| | TCop-Kmeans [176] (2021) | individual opinions, trust ratings from other users | social network analysis, clustering, similarity measures | leverage trust relationships as weights to develop a minimum-cost consensus model; facilitate consensus reaching in large-scale decision-making |
| **8. Decentralized TM** | Social-Chain [173] (2021) | social behaviors, communication interactions on the blockchain | public-private keys, consensus mechanisms, timestamp validation algorithms | proposed a robust blockchain-based decentralized trust evaluation model; introduced a new consensus mechanism named Proof-of-Trust (PoT) |
| | Yu et al. [175] (2023) | transactions and messages on the blockchain | smart contracts, sentiment analysis, RoBERTa [102], IPFS [7] | proposed a distributed reputation mechanism to distinguish the authenticity of information and prevent the spread of misinformation. |

| | Models | Data Requirements | Algorithms/Techniques | Application Scenarios |
|---|---|---|---|---|
| | Sarode et al. [133] (2023) | n/a. | Ethereum blockchain, smart contracts, IPFS [7], SHA-256 | introduced a novel decentralized social network; incentive mechanisms reward high-quality content contributions and promote reliable participation |
| 9. Cognitive TM | CHASSIS [90] (2020) | user online social activities (tweets, retweets, posts, comments, likes) | conformity [3, 26, 31], diffusion trees [89], Hawkes process [57, 185] | introduced a novel information diffusion model based on a social psychology theory named conformity; investigate the interplay between informational conformity and normative conformity |
| | Karami et al. [76] (2021) | user profile, number of followers/friends, user online social activities (tweets, retweets, posts) | rumor psychology [33], words psychology [147], LIWC features [15, 124], BERT [32] | to analyze the characteristics and motivations of individuals who spread false news on social media |
| | CPTT [39] (2020) | n/a. | predictive processing in cognitive theory [27, 155] | depict how individuals rely on cognitive predictions to form trust judgments during interactions and information exchanges |
| 10. Multi-Component TM | TrustLab™ [61, 62, 179] (2021) | number of followers, user online social activities (tweets, retweets, posts), user characteristics | sentiment analysis, PageRank [120], HITS [79], Random Forests | measure user trustworthiness on social media with multiple trust attributes; visualize the distribution of Twitter (X) user trust scores |
| | iSim [42] (2021) | trust ratings of users to their peers | vector space similarity, time-aware matrix factorization, trust propagation, similarity measures | proposed a multi-component trust prediction method based on user similarity |
| | T-VLC [20] (2023) | participant characteristics, interactions between participants | Bayesian networks, social network analysis, centrality measures, Pearson coefficient | proposed a general trust model for assessing trust among participants in online learning communities |

## 7 HANDBOOK FOR TRUST MODELING

As George Box stated, "all models are wrong, but some models are useful" [13]. In this section, we will provide an implementation-centric trust modeling handbook to inspire the development and implementation of trust models. Based on the literature review, we summarize the available datasets, trust-related features, promising modeling techniques, and feasible application scenarios. Overall, the effectiveness of trust modeling depends on the quality of the data, the relevance of the trust features, and the suitability of the chosen algorithm or approach.

### 7.1 Available Data to Compute Online Trustworthiness

In this section, we attempt to address the question: What information or data is available to compute online trustworthiness? We explored various variables and datasets from online social networks that may indicate online trustworthiness. Table 2 summarizes the datasets available for trust modeling and experimental evaluation. Relevant and diverse datasets provide the foundation for making reasonable assumptions in trust modeling. Selecting features that are highly correlated with trust and can effectively characterize trust is also crucial. In Figure 5, we outline four types of datasets commonly used to calculate online trust. For each type of dataset, we identified several trust-related features that can be used to build trust models.

Table 2. Available Datasets for Trust Modeling in Online Social Networks.

| Datasets | Source Links | Details of the Datasets | User Profile | Trust Relationships | Connections | Interactions | Reviews, Ratings |
|---|---|---|---|---|---|---|---|
| Advogato [109] | https://networkrepository.com/advogato.php | 6,541 users and 51,127 trust relationships. Four trust relationships {Observer, Apprentice, Journeyer, Master} | | ✓ | | | |
| Pretty Good Privacy (PGP) [11] | https://networkrepository.com/tech-pgp.php | 38,546 users and 317,979 trust relationships. Four levels of trust relationships. | | ✓ | | | |
| BlogCatalog [146] | http://datasets.syr.edu/datasets/BlogCatalog3.html | 10,312 bloggers (nodes), 39 groups and 333,983 friendship pairs (binary, undirected edges). | ✓ | | ✓ | | |
| Flickr [146] | http://datasets.syr.edu/datasets/Flickr.html | 80,513 users (nodes), 195 groups and 5,899,882 friendship pairs (binary, undirected edges). | ✓ | | ✓ | | |
| Youtube [146] | http://datasets.syr.edu/datasets/YouTube2.html | 1,138,499 users (nodes), 47 groups and 2,990,443 friendship pairs (binary, undirected edges). | ✓ | | ✓ | | |
| Social Circles: Facebook [86] | https://snap.stanford.edu/data/ego-Facebook.html | 4,039 users (nodes) and 88,234 undirected edges. | ✓ | | ✓ | | |
| Social Circles: Twitter [86] | https://snap.stanford.edu/data/ego-Twitter.html | 81,306 users (nodes) and 1,768,149 directed edges. | ✓ | | ✓ | | |
| Social Circles: Google+ [86] | https://snap.stanford.edu/data/ego-Gplus.html | 107,614 users (nodes) and 13,673,453 directed edges. | ✓ | | ✓ | | |
| DBLP [145] | https://www.aminer.org/citation | 5,259,858 papers and 36,630,661 citation relationships. | ✓ | | ✓ | | |
| Cora [136] | https://linqs.org/datasets/#cora | 2,708 scientific publications, 7 classification classes and 5,429 links (binary, directed edges). | ✓ | | ✓ | | |
| Citeseer [130] | https://linqs.org/datasets/#citeseer-doc-classification | 3,312 scientific publications, 6 classification classes and 4,732 links (binary, directed edges). | ✓ | | ✓ | | |
| PubMed | https://linqs.org/datasets/#pubmed-diabetes | 19,717 scientific publications and 44,338 links (binary, directed edges). | ✓ | | ✓ | | |
| Cresci-17 [28] | http://mib.projects.iit.cnr.it/dataset.html | 3,474 legitimate users, 10,894 fake bots, 6,637,616 tweets and relations. | ✓ | | ✓ | ✓ | |
| Midterm-18 [174] | https://botometer.osome.iu.edu/bot-repository/datasets.html | 8,092 legitimate users and 42,446 fake bots. | ✓ | | ✓ | ✓ | |
| TwiBot-22 [40] | https://github.com/LuoUndergradXJTU/TwiBot-22 | 860,057 legitimate users, 139,943 fake bots, 86,764,167 tweets and 170,185,937 relations. | ✓ | | ✓ | ✓ | |
| MGTAB [140] | https://github.com/GraphDetec/MGTAB | 7,554 legitimate users, 2,830 fake bots and 7 types of relations. | ✓ | | ✓ | ✓ | |
| Ciao [144] | https://www.cse.msu.edu/~tangjili/trust.html | 7,317 users, 283,319 ratings and 111,781 trust relations between users. | ✓ | ✓ | | | ✓ |
| Epinions [144] | https://www.cse.msu.edu/~tangjili/trust.html | 22,164 users, 922,267 ratings and 355,754 trust relations between users. | ✓ | ✓ | | | ✓ |

### 7.1.1 Connections, Relationships.

Trust can be inferred from user connections in the social network graph. Users in the same community tend to exhibit similar behaviors and levels of trust, and users who connect with trusted users are also likely to be considered more trustworthy. Trust features related to user connections or relationships include:

- **Node Connectivity**: number, strength, and type of connections (e.g., friends, followers, or mutual connections)
- **Centrality Measures**: degree centrality, betweenness centrality, eigenvector centrality, and closeness centrality
- **Community Detection**: inferring trust levels by identifying clusters or communities in social network graphs

### 7.1.2 Behaviors, Interactions, Transactions.

Trust relationships between users can be evaluated by analyzing their past behaviors, interactions, or transactions. Consistency in behavior can indicate trustworthiness, while inconsistency or sudden changes in behavior may indicate untrustworthiness. The frequency and type of interactions between users can also indicate trust levels. Generally speaking, trust can be measured by the following features:

- **Interaction Patterns**: frequency, consistency, and types of interactions (e.g., likes, shares, comments, messages, or group chats)
- **Behavior Patterns**: frequency, regularity, and timing of online behavior or social activities
- **Bot or Anomaly Detection**: the presence of automated, bot-like activities or abnormal behavior patterns

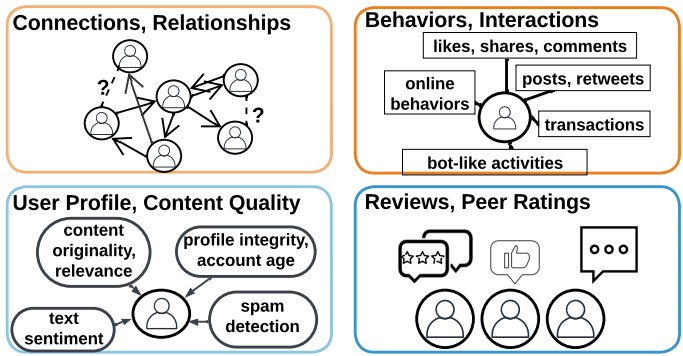

Fig. 5. Four commonly used datasets for calculating online trust.

### 7.1.3 User Characteristics, Content Quality.

Analyzing user characteristics and user-generated content can also provide insights into user trust levels. Verified profiles and long-registered accounts are generally considered more trustworthy. High-quality, informative user-generated content is more likely to gain trust [6]. Representative trust features include:

- **User Profile**: profile integrity, account age, and consistency of personal information (i.e. IP address, name, bio, and profile picture)
- **Content Quality**: spelling, grammar, originality, coherence, informativeness, and relevance of content
- **Text Sentiment**: the sentiment of text content (e.g., positive, negative, or neutral), the emotional tone of user-generated content
- **Spam Detection**: the presence of spam, duplication, misinformation, disinformation, or manipulated content

### 7.1.4 Reviews, Ratings.

Trust models can also leverage ratings and reviews provided by users to evaluate user trustworthiness. There are two types of rating data. The first comes from online review platforms, where users provide feedback on the quality of services and products. The second type involves trust ratings, which users assign to their peers within the community. Trust features related to user-provided ratings or reviews include:

- **Product Review or Rating Behaviors**: rating consistency, rating distribution, review length, review frequency
- **Peer Endorsements**: reputation scores accumulated through interactions, aggregated peer ratings, endorsements, or badges

## 7.2 Promising Techniques for Building Trust Models

Various techniques and algorithms have been explored to build trust models. This section focuses on the research question: How is trust quantified? What algorithms can be used to build trust quantification models? In Table 3, we outline several representative trust modeling techniques, highlighting the unique perspective each technique brings to the field of trust modeling.

Table 3. Promising Techniques for Building Trust Models in Online Social Networks.

| Techniques | Trust Modeling Mechanisms and Challenges | Instances |
|---|---|---|
| Reputation-Based | • User trustworthiness is measured through a reputation score, which is accumulated through behavior, contributions, and peer feedback during interactions. It is computationally transparent and widely applicable.<br>∗ The challenge is that reputation trust models are vulnerable to malicious manipulation and Sybil attacks. | Rezvani and Rezvani [128] (2020), GOLI [65] (2023) |
| Probabilistic /Bayesian | • Represent trust as a probability distribution and update trust using Bayesian inference or Beta distribution when new evidence emerges. They work well with sparse interactions or noisy data.<br>∗ The problem is that modeling requires prior probabilities, and it is difficult to set prior probabilities without historical data. | Lingam et al. [94] (2020), Naumzik and Feuerriegel [117] (2022) |
| Hidden Markov Models (HMMs) | • Represent user trust as a hidden state (e.g., "low trust" or "building trust"). Transitions between trust states are computed based on the observed interactions using transition probabilities. It is suited to model sequences of hidden states and capture the temporal dynamics of trust.<br>∗ The challenge is that HMMs require sufficient historical data to estimate the parameters. | ElSalamouny et al. [35] (2009), Liu and Datta [101] (2012) |
| Subjective Logic | • Represent trust relation as a combination of beliefs and uncertainties [belief $b$, disbelief $d$, uncertainty $u$], $(b, d, u \in [0, 1])$. Dempster-Shafer theory [80, 137, 172] can be used to combine the three parameters and compute a final trust value. It models the subjectivity and uncertainty inherent in trust relationships.<br>∗ The definitions of opinion triplets and membership functions can vary in different contexts and may contain biases. | Cho et al. [24] (2019), Liu et al. [98] (2021) |
| Fuzzy Logic | • Represent trust as a linguistic variable with fuzzy sets (e.g., "high trust" or "medium trust"). It deals with ambiguity in trust and is valuable when trust cannot be modeled numerically. Additionally, multiple trust factors can be aggregated using operators such as "AND", "OR", and "NOT".<br>∗ The difficulty lies in designing appropriate membership functions [58, 121] and fuzzy rules. | FTWDF-EEFA [104] (2021), Ma et al. [105] (2021) |
| Sentiment Analysis | • Inferring user trustworthiness by evaluating the emotional tone of user-generated content [95, 122]. Positive sentiment indicates trust, while negative sentiment indicates distrust.<br>∗ However, these models often simplify emotions into binary or ternary categories, which may overlook nuanced emotions related to trust. Moreover, sentiment analysis often struggles to detect sarcasm or irony. | SQUAD [85] (2024), DDTrust [45] (2020), TextBlob [103] (2018) |
| Trust Game | • User interactions are modeled as repeated trust games, where individuals decide whether to trust or deceive based on potential rewards and risks [70, 143]. It is well-suited for market-oriented social networks where incentives matter.<br>∗ Problems arise when users use pseudonyms or frequently change accounts, making it difficult to enforce long-term strategies or penalties for breaches of trust. | Hu et al. [59] (2021), Guo et al. [53] (2022) |

| Techniques | Trust Modeling Mechanisms and Challenges | Instances |
|---|---|---|
| Social Network Analysis (SNA) | • Represent the social network as a weighted graph, users as nodes, and trust relationships as weighted edges. Infer trust between users based on the network structure and connectivity. Common methods include graph theory and centrality measures (e.g., degree centrality, betweenness centrality, eigenvector centrality, and closeness centrality).<br>∗ One challenge is that the model is vulnerable to malicious link attacks, and another challenge is the high computational cost in large-scale social networks. | Meo et al. [114] (2017), T-VLC [20] (2023), INTRUST [170] (2025) |
| Trust Propagation | • Infer trust relationships from connections and paths. Graph traversal algorithms such as breadth-first search and depth-first search can be used to build a complete trust network based on existing relationships.<br>∗ Trust dilution along long paths. Moreover, trust evaluations may conflict when they are propagated from multiple sources, thereby requiring an appropriate aggregation mechanism. | AHNTP [169] (2024), Wu et al. [165] (2021), 3VSL-AT [98] (2021) |
| Graph Neural Networks (GNNs) | • Users are nodes, and each node aggregates trust-related signals from its neighbors to update its representation. After several layers of message passing, the trust score is predicted based on the final representation of the node. It can capture complex relational dependencies and is widely used in trust prediction and trust propagation.<br>∗ Nevertheless, the complex structure of GNNs leads to limited model interpretability, and their performance is further limited by data sparsity. | TrustGNN [63] (2023), Gatrust [68] (2022), Trustguard [157] (2024) |
| Supervised Learning | • It leverages historical data to predict user trustworthiness in new interactions. First, it extracts relevant features from historical data, such as user behavior, interaction patterns, or text sentiment. Next, it uses supervised algorithms (e.g., SVMs, random forests, or neural networks) to learn the mapping between these features and labeled trust levels.<br>∗ Yet, it requires a large amount of labeled data (trusted/untrusted) for training. | NeuralWalk [96] (2019), Guardian [93] (2020) |
| Community Detection | • Group users into communities based on similar trust behaviors or interaction patterns. It can detect hidden trust relationships between users, especially in the absence of explicit trust labels. It also helps identify and isolate malicious users or groups.<br>∗ Applying community detection to large-scale social networks can be computationally expensive. | UnDBot [123] (2024), SBCD [94] (2020) |
| Anomaly Detection | • Identify unusual patterns in user behavior that deviate from normal. It establishes a baseline of anticipated activity to flag potential fraudulent behavior or security threats. Common techniques include statistical methods (e.g., Z-scores, 3-sigma rule) or machine learning methods (e.g., isolation forests, graph neural networks).<br>∗ User behavior and interactions are evolving (concept drift). Therefore, adaptive or online learning models are needed to continuously update the trust or anomaly thresholds. | TA-Detector [162] (2024), THGNN [92] (2023) |
| Blockchain | • Blockchain enables decentralized trust computation, immutability of interactions, and transparency in trust relationships. Cryptographic protocols and consensus mechanisms can be used for trust evaluation. Smart contracts can automatically calculate and update trust scores.Tokenization and incentive mechanisms help promote honest and trustworthy behavior.<br>∗ This approach is still in its early stages but holds potential. | Social-Chain [173] (2021), Yu et al. [175] (2023) |
| Cognitive Models | • Incorporate principles from psychology and cognitive science into trust models, as well as human factors such as emotional bonding, cognitive assessment, and behavioral analysis. | Karami et al. [76] (2021), CHASSIS [90] (2020) |

Modern trust modeling techniques are built upon basic trust models. Next, we will explore several basic trust models and discuss the mechanisms they use to quantify trust in online social networks. We also provide the latest variants of these base models for further reference.

### 7.2.1 *Reputation Trust Model*.

Trust is inferred from the reputation score accumulated through multiple interactions. The reputation score of user $i$ is calculated by aggregating the feedback provided by its peers, $R_i = \frac{\sum_{j \in S_i} t_{ij} \times \omega_j}{\sum_{j \in S_i} \omega_j}$, where $S_i$ is the set of users that provide peer feedback about user $i$, $t_{ij}$ is the feedback from user $j$ to user $i$, and $\omega_j$ is the weight of the feedback from user $j$.

For example, the Beta reputation model [74] utilizes the Beta probability distribution to calculate trustworthiness. The trust evaluation value of $A$ towards $B$ is calculated as $T_{A,B} = \frac{\alpha+1}{\alpha+\beta+2}$, where $\alpha$ is the number of positive interactions and $\beta$ is the number of negative interactions. Furthermore, if multiple users $C_1, C_2, \cdots, C_n$ have interacted with $B$ and provided trust evaluations, $A$ can aggregate these trust evaluations to form an overall trust evaluation of $B$, $T_{A,B} = \sum_{i=1}^{n} T_{C_i,B} \times \omega_i$, where $\omega_i$ is the weight of the trust evaluation provided by $C_i$.

Reputation and trust models are commonly used in e-commerce and community-based applications, where peer feedback is crucial. State-of-the-art variants of reputation trust models include Rezvani and Rezvani [128] (2020) and GOLI [65] (2023).

### 7.2.2 *EigenTrust Model*.

Online social networks can be represented as weighted graphs, where users are nodes and trust relationships are weighted edges. EigenTrust [75] calculates a global trust score for each node by aggregating local trust values rated by peer nodes (Algorithm 1).

---

**Algorithm 1:** EigenTrust [75]

---

1 **Local Trust Values:** Nodes rate their peer nodes based on past interactions.
2 **Normalization and Trust Matrix:** Local trust values are normalized to ensure they sum to 1 for each node. $c_{ij}$ is the local trust value that node $i$ assigns to node $j$ and normalized to $s_{ij} = \frac{c_{ij}}{\sum_k c_{ik}}$. The normalized local trust values form the trust matrix $\mathbf{S}$, where $\mathbf{S}_{ij} = s_{ij}$.
3 **Global Trust Computation:** Compute the principal eigenvector of the trust matrix $\mathbf{S}$ using the power iteration method and obtain the global trust value vector $\mathbf{t}$.
4 **Iterative Computation:** The global trust vector $\mathbf{t}$ is computed iteratively using Equation 1.

---

The global trust vector $\mathbf{t}$ is the solution to the equation, $\mathbf{t} = \mathbf{S}^T \mathbf{t}$, subject to the normalization condition $\sum_i t_i = 1$ and $t_i \geq 0$ for all $i$, where $t_i$ is the global trust value of node $i$. The algorithm iteratively updates the trust vector $\mathbf{t}$ until it converges,

$$\mathbf{t}^{(k+1)} = d\,\mathbf{S}^T\,\mathbf{t}^{(k)} + (1-d)\,\mathbf{t}^{(0)} \tag{1}$$

where $d$ is the damping factor (typically around 0.85) [18] and $\mathbf{t}^{(0)}$ is the initial trust vector (usually set to a uniform distribution $\mathbf{e}$).

One of the latest variants of the EigenTrust model is Vedula et al. [152] (2017).

### 7.2.3 *Trust Propagation Model*.

According to the transitivity of trust [51], if $A$ trusts $B$, and $B$ trusts $C$, then $A$ may also trust $C$ to some extent.

$$T_{A,C} = T_{A,B} \times T_{B,C} \tag{2}$$

where $T_{A,B}$ represents the trust value of $A$ towards $B$ , $T_{B,C}$ represents the trust value of $B$ towards $C$, and $T_{A,C}$ represents the trust value of $A$ towards $C$, which is derived from $T_{A,B}$ and $T_{B,C}$.

State-of-the-art variants of trust propagation model include Wu et al. [165] (2021), AtNE-Trust [159] (2020), and 3VSL-AT [98] (2021).

### 7.2.4 *TrustRank Model*.

TrustRank is a link-analysis algorithm based on the PageRank algorithm [120], designed to combat web spam and low-quality content [55]. The core idea is that high-quality web pages will link to other high-quality web pages, while spam pages rarely receive links from trusted pages.

First, construct a web graph where web pages are nodes and hyperlinks are directed edges. TrustRank starts with a small set of manually verified trusted pages (trusted seeds) and propagates trust through outgoing links (Algorithm 2).

---

**Algorithm 2:** TrustRank [55]

---

1 **Selection of Seed Set:** Manually select a small set of highly trusted pages (nodes) as trusted seeds $S$.
2 **Trust Vector Initialization:** $T_0$ is the initial trust vector. For each node in the seed set $n \in S$, $T_0(n) = 1$, otherwise $T_0(n) = 0$ .
3 **Iterative Computation:** Iteratively update the trust vector $\mathbf{T}$ until trust scores stabilize (Equation 3).

---

The transition probability matrix is $\mathbf{P}$, where $P_{mn}$ represents the probability of transitioning from node $m$ to node $n$. If there is a link from $m$ to $n$, then $P_{mn} = \frac{1}{Out(m)}$, where $Out(m)$ is the number of outgoing links from $m$. The trust vector $\mathbf{T}$ is iteratively calculated using the equation,

$$\mathbf{T}^{(k+1)} = (1-d)\,\mathbf{T}_0 + d\,P^T\,\mathbf{T}^{(k)} \tag{3}$$

where $d$ is the damping factor (typically around 0.85) [18]. The algorithm converges when the change in trust values between two consecutive iterations falls below a small threshold, $\left\| \mathbf{T}^{(k+1)} - \mathbf{T}^{(k)} \right\| < \epsilon$. Generally, pages that are closer to trusted seeds (calculated by link distance) have higher trust scores.

State-of-the-art variants of PageRank-based trust models include MPR [184] and AHNTP [169].

### 7.2.5 *Subjective Logic Trust Model*.

Subjective logic is a probabilistic logic framework that explicitly models uncertainty when expressing trust [22, 72, 73]. Trust is represented by an opinion tuple $\omega = (b, d, u, a)$, and $b + d + u = 1$. Belief ($b$) indicates the degree to which a proposition is believed to be true, disbelief ($d$) indicates the degree to which it is believed to be false, uncertainty($u$) reflects the degree of uncertainty about its truth, and base rate ($a$) is the prior probability that the proposition is true without any evidence. Subjective logic has several operators for trust [73]:

- **Discounting Operator**: When A trust B with opinion $T_{A,B} = (b_{AB}, d_{AB}, u_{AB}, a_{AB})$, and B provides an opinion about the proposition $x$, $\omega_{B \to x} = (b_x, d_x, u_x, a_x)$. How much does A trust the proposition $x$?

$$\omega_{A \to x} = T_{A,B} \otimes \omega_{B \to x} = (b_{AB} \cdot b_x, b_{AB} \cdot d_x, d_{AB} + u_{AB} + b_{AB} \cdot u_x, a_x) \tag{4}$$

  where $\omega_{A \to x}$ is the opinion of A about the proposition $x$ after discounting the opinion of B about the proposition $x$. Discounting is applying A's trust in B to B's opinion.

- **Consensus Operator (Fusing Opinions)**: Aggregate multiple independent opinions. Assume that $A$ and $C$ both have an opinion about whether B is trustworthy ($\omega_{A,B}$ and $\omega_{C,B}$). Then the aggregated opinion is,

$$\omega_{A,B} \oplus \omega_{C,B} = \left( \frac{b_{A,B} \cdot u_{C,B} + b_{C,B} \cdot u_{A,B}}{u_{A,B} + u_{C,B} - u_{A,B} \cdot u_{C,B}}, \frac{d_{A,B} \cdot u_{C,B} + d_{C,B} \cdot u_{A,B}}{u_{A,B} + u_{C,B} - u_{A,B} \cdot u_{C,B}}, \frac{u_{A,B} \cdot u_{C,B}}{u_{A,B} + u_{C,B} - u_{A,B} \cdot u_{C,B}}, a_{A,B} \right) \tag{5}$$

  where the base rate is assumed to be the same for both opinions, $a_{A,B} = a_{C,B}$.

State-of-the-art variants of subjective logic trust models include Cho et al. [24] (2019) and Liu et al. [98] (2021).

### 7.2.6 *Fuzzy Trust Model*.

Fuzzy logic deals with the ambiguity in trust using linguistic terms (e.g., "high trust" or "low trust") and membership functions [177, 178]. There are several key components:

(1) **Fuzzy Sets and Membership Functions**: Suppose that a seller's trustworthiness is evaluated based on two factors: the number of positive reviews (Experience) and the average rating (Reputation). The input and output fuzzy sets can be defined as:
   - Input - Experience: {Low, Medium, High}, Reputation: {Poor, Average, Good}
   - Output - Trust: {Low, Medium, High}

   Membership functions convert crisp inputs into fuzzy values [58, 121]. Here is an example of a triangular membership function for "Low Experience": $\mu_{E_L}(x) = \begin{cases} 1 & \text{if } x \leq a \\ \frac{b-x}{a} & \text{if } a < x \leq b \\ 0 & \text{if } x > b \end{cases}$, where $x$ is the number of positive reviews, $a$ and $b$ are conditional thresholds, and $\mu_{E_L}(x) \in [0, 1]$. The membership functions for "Medium Experience" ($\mu_{E_M}(x)$) and "High Experience" ($\mu_{E_H}(x)$) are similar, except that the threshold conditions are different.

(2) **Fuzzy Rules**: Define a set of fuzzy rules to map input fuzzy variables to output fuzzy variables.
   **Rule 1:** IF Experience is High AND Reputation is Good, THEN Trust is High.
   **Rule 2:** IF Experience is Medium AND Reputation is Average, THEN Trust is Medium.
   **Rule 3:** IF Experience is Low OR Reputation is Poor, THEN Trust is Low.

(3) **Fuzzy Inference**: Apply fuzzy rules to calculate fuzzy output variables. For example, applying **Rule 1** using the Mamdani inference method: IF $\mu_{E_H}(x)$ AND $\mu_{R_G}(y)$ THEN $\mu_{T_H}(z) = min(\mu_{E_H}(x), \mu_{R_G}(y))$.

(4) **Defuzzification**: Combine the fuzzy outputs of all rules to form a single fuzzy set and convert the aggregated fuzzy set $\mu_T(z)$ into a single crisp trust value $t_0$. For example, using the centroid method, $t_0 = \frac{\int z \cdot \mu_T(z) \, dz}{\int \mu_T(z) \, dz}$.

State-of-the-art variants of fuzzy trust models include FTWDF-EEFA [104] (2021) and Ma et al. [105] (2021).

### 7.2.7 *Bayesian Trust Model*.

Bayesian trust models use Bayesian inference [14] to update the probability of an user being trustworthy based on new evidence or interactions [25, 161]:

$$P(T_{A,B}^{\text{posterior}} \mid E) = \frac{P(E|T_{A,B}) \times P(T_{A,B}^{\text{prior}})}{P(E)} \tag{6}$$

where $T_{A,B}$ represents the degree of trust that A has in B, $P(T_{A,B}^{prior})$ is the prior probability, the initial trust value before any interaction or evidence, $P(E)$ is the probability of observing evidence $E$, $P(E|T_{A,B})$ is the probability of observing evidence $E$ given the current trust value $T_{A,B}$, and $P(T_{A,B}^{posterior} \mid E)$ is the posterior probability of trust given evidence $E$.

State-of-the-art variants of Bayesian trust models include Cho et al. [24] (2019), Lingam et al. [94], Naumzik and Feuerriegel [117] (2022), 3VSL-AT [98] (2021), and T-VLC [20] (2023).

### 7.2.8 *Game Theoretic Trust Model*.

Game theory models trust as strategic interactions between rational players [70, 143]. Players evaluate whether to trust or deceive each other based on the potential rewards and risks in the trust game.

A well-known model is the repeated Prisoner's Dilemma [82, 125, 126]. The payoff matrix of a single-round Prisoner's Dilemma is:

| Player A \ Player B | Trust | Defect |
|---|---|---|
| Trust | $(R, R)$ | $(S, T)$ |
| Defect | $(T, S)$ | $(P, P)$ |

where $R$ is the reward for mutual trust, $S$ is the payoff of the player who chose to trust and was deceived, $T$ is the temptation of deception, and $P$ is the punishment for mutual deception. Typically, $T > R > P > S$.

Suppose player A trusts player B with probability $T_{AB}$, and the expected payoff of player A is $E_A$. If A chooses to cooperate, then $E_A(C) = T_{AB} \cdot R + (1 - T_{AB}) \cdot S$, otherwise, if A chooses to defect, then $E_A(D) = T_{AB} \cdot T + (1 - T_{AB}) \cdot P$. In the repeated trust game, A can adjust trust $T_{AB}$ according to the difference between actual payoff and expected payoff:

$$T_{AB}^{(t+1)} = T_{AB}^{(t)} + \alpha(R - E_A) \tag{7}$$

where $\alpha$ is a learning rate.

State-of-the-art variants of game-theoretic trust models include PRADA-TF [106] (2024), Hu et al. [59] (2021), and Guo et al. [53] (2022).

### 7.3 What Application Scenarios Benefit from Trust Models

This section revolves around the question: What application scenarios benefit from trust quantification models? Below, we outline eight application scenarios of online social networks where trust models play an important role.

#### 7.3.1 Building Connections and Community Formation.

An important role of trust models in online social networks is to measure, infer, and manage trust relationships between users.

(1) **Building Connections:** Suggesting trustworthy users to follow or connect with based on past interactions, shared connections, and user reliability [36].
(2) **Community Formation:** Facilitating the formation of trust-based communities or discussion groups where users are more likely to engage productively.

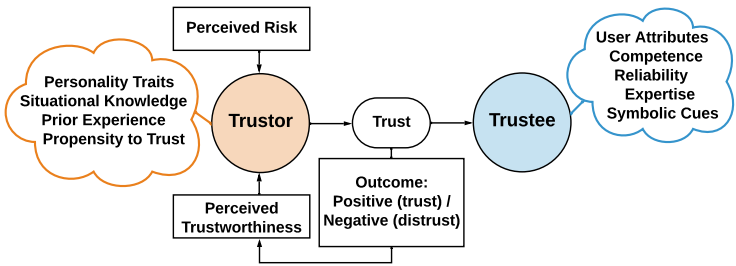

Fig. 6. Trust Formation Model [9, 10, 110]

Figure 6 illustrates the formation of trust between a trustor and a trustee [9, 10]. From the trustee's perspective, influencing factors include user attributes (e.g., age, gender, education, residence, income, and ethical standards), competence, reliability, expertise, and symbolic cues that indicate trust. From the trustor's perspective, influencing factors include personality traits, situational knowledge, prior experience, and perceived trustworthiness.

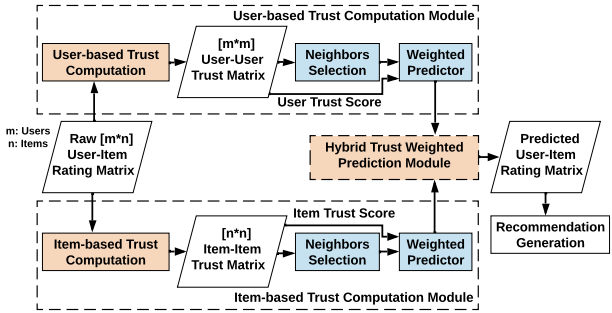

Fig. 7. Trust-Based Collaborative Filtering [138]

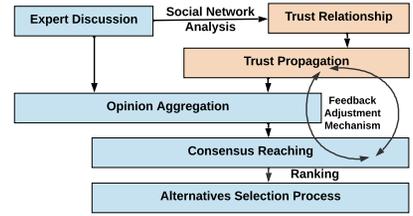

Fig. 8. Trust-Driven Group Decision-Making [151, 181]

### 7.3.2 *Information Credibility and Fake News Detection*.

Trust models assign trust scores to users based on their past behavior, expertise, and reputation. The credibility of information can be inferred from the trustworthiness of its source [2].

(1) **Information Credibility Assessment**: Combining content features (e.g., linguistic style, sentiment, and fact-checking consistency) with trust features (e.g., user reliability and trusted endorsements) to measure information credibility. It helps filter out misleading information that appears linguistically credible but originates from low-trust accounts.

(2) **Fake News Detection**: Combating fake news, rumors, harmful propaganda, and bot-driven campaigns [182].

### 7.3.3 *Content Filtering and Recommendation*.

Trust models distinguish and promote posts, articles, or media shared by trusted users when personalizing information feeds. Furthermore, trust models improve the accuracy of recommendation methods compared to those based solely on popularity or similarity.

(1) **Content Filtering**: Filtering and ranking posts, articles, or videos based on the trustworthiness of the source.

(2) **Product Recommendations**: Recommending products or services based on the preferences of trusted peers [188].

Trust-based collaborative filtering improves recommendation performance by prioritizing the opinions of trusted neighbors. As shown in Figure 7, the calculated trust scores are first applied to rank and select trusted neighbors and then used as weighting factors when aggregating their preferences. Trust models also help alleviate the data sparsity and cold start problems inherent in collaborative filtering [60].

### 7.3.4 *Group Decision-Making and Consensus Reaching*.

In scenarios where users collectively make decisions, trust models help prioritize and promote opinions from reliable sources. This filtered view enables more efficient decision-making and allows the group to focus on information from highly trusted sources.

As illustrated in Figure 8, group decision-making includes four main steps: expert discussion, opinion aggregation, consensus reaching, and alternative selection. Trust, as a motivating and weighting factor, is incorporated into the feedback and adjustment process and accelerates consensus reaching [100, 151, 181].

### 7.3.5   *Online Collaboration and Crowdsourcing*.

Trust models help identify reliable contributors based on past contributions, collaboration history, and peer reviews. Through trust scores, collaboration and crowdsourcing platforms can prioritize tasks to reliable contributors, thereby facilitating online collaboration and knowledge sharing.

### 7.3.6   *Social Commerce and Transactions*.

In social platforms with market functions, trust models help evaluate the reliability of sellers and enhance transaction confidence [150].

(1) **Seller and Buyer Trust:** Building trust between buyers and sellers to enable peer-to-peer transactions [107].
(2) **Secure Transactions:** Identifying fake or biased reviews and preventing fraud in transactions.

### 7.3.7   *Influence Analysis and Information Diffusion*.

Trust models can enhance the identification of influencers beyond just the number of followers. Selecting influencers based on trustworthiness can help correct or counteract misleading content, reduce the spread of rumors, and avoid the pitfalls of misinformation [91].

(1) **Influencer Identification:** Ranking individuals by trust to identify authoritative voices or opinion leaders in communities [1, 65, 108].
(2) **Information Diffusion:** Trust-aware information diffusion models help identify whether trending content is likely to be misinformation.

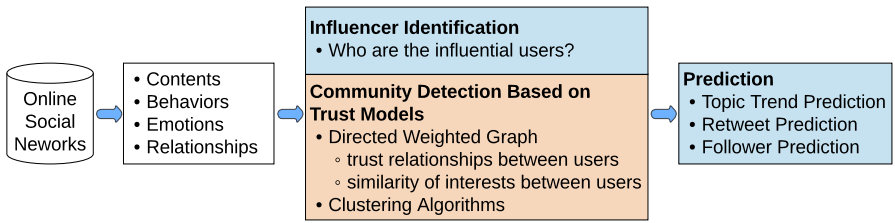

Fig. 9. Trust-Based Community Detection and Information Diffusion [91, 149]

Figure 9 shows a trust-based community detection and information diffusion model [91, 149]. Trust models were used for influencer identification and community detection, enhancing the accuracy of information diffusion prediction.

### 7.3.8   *Security and Privacy*.

Trust models help prevent fraud and support privacy protection, strengthening the security of online interactions [129].

(1) **Fraud Detection:** Identifying malicious, fake, or sybil accounts and detecting potential fraud, phishing, or spam by evaluating unusual behavior patterns [94, 118, 123].
(2) **Access Control:** Granting or restricting access to personal or sensitive information based on trust levels.

## 8   DISCUSSION AND FINDINGS

**To answer the question, "What is trust? How is trust defined in online social networks,"**
Trust is a willingness to accept vulnerability based on positive expectations of others. It stems from both rational evaluation and emotional bonds. Trust takes time to build, especially in anonymous online communities, but it can also be destroyed quickly. In online social network, trust can be defined as the confidence that users have in the reliability, integrity, and security of social networking platforms and their fellow users:

- Reliability: Confidence in the predictable behavior of fellow users and the consistency and reliability of online interactions.
- Integrity: Confidence in the authenticity of shared content, explainability of decision-making algorithms, and transparency in data use.
- Competence: Confidence in the capabilities of online platforms and the security against unauthorized access or cyber threats.

**To answer the question, "What data is available to compute online trustworthiness?",** computing online trustworthiness involves analyzing various datasets and carefully selecting trust features to formulate effective trust models. Considering availability and practicality, we outline four dataset categories and their corresponding trust-related features. The most commonly used trust features are: (1) the number and strength of connections, (2) the frequency and consistency of interactions, (3) the quality and relevance of shared content, and (4) reputation scores established through peer review or endorsement. In short, ensuring data quality and selecting the most relevant trust features are key to effective trust modeling in online social networks.

**To answer the question, "How is online trust quantified?",** we categorize the latest trust quantification models into ten algorithmic categories. In reputation-based trust models, trust is quantified as a numerical score accumulated and aggregated from past behaviors, including interactions, ratings, or endorsements from peers. Probabilistic trust models represent trust as a probability distribution and utilize Bayesian inference to update posterior trust beliefs with new evidence [83, 161]. Subjective logic and fuzzy logic deal with uncertainty and ambiguity in trust relationships [72, 121]. Fuzzy logic models express trust levels using linguistic terms such as "low", "medium", and "high" and update them according to fuzzy rules [177, 178]. Moreover, graph-based trust models treat the social network as a graph where nodes are users and edges carry trust scores [69]. Trust is inferred from connections and can be aggregated along multiple paths (e.g., weighted average of neighbors' trust). Machine learning models exploit features of interactions and network structure to learn trustworthiness and can model nonlinear patterns [156]. For example, graph neural networks learn trust embeddings directly from social graphs. In addition, game theoretic models model trust as the outcome of strategic interactions. Hybrid models integrate social graph features, peer ratings, and behavioral data into one trust score. Although these trust models originate from different modeling perspectives, they collectively deepen our understanding of trust and transform trust from an abstract concept into a concrete numerical variable in online social networks.

**To answer the question, "What application scenarios benefit from trust quantification models?",** three have been identified. First, trust models help form a vibrant and supportive community. Trust models help foster trust-based communities or discussion groups, where users are more likely to engage in productive and effective interactions. Additionally, trust models can identify reliable and influential users who can act as moderators to guide community interactions and encourage the dissemination of accurate, high-quality content. Second, trust models help maintain content quality and support group decision-making and crowdsourcing. Trust can be used as a screening criterion to filter out misleading or harmful content, while prioritizing reliable content [141, 187]. Moreover, trust scores can be used as weights in opinion aggregation or crowdsourcing, facilitating collaboration or group decision-making. Third, trust models can enhance the security of online social networks. They help detect fake profiles and malicious users, thereby reducing the risk of fraud or deception. By analyzing user behavior patterns and social connections, trust models can identify and prevent suspicious activities or fraudulent behavior early on.

## 9 OPEN CHALLENGES AND RESEARCH DIRECTIONS

Trust modeling in online social networks is challenging due to data sparsity, model bias, evolving trust, privacy issues, and malicious manipulation. In this section, we highlight unresolved challenges and suggest potential directions for progress.

### 9.1 Data Sparsity and Model Bias

Trust often has to be inferred from partial or noisy data, as trust information is rarely complete due to missing profiles, private interactions, etc. Also, trust modeling suffer from data sparsity and cold start problems. It is difficult to calculate a reliable trust score for new users or users with limited interactions. Moreover, bias and unfairness in trust models [134] may unfairly favor or penalize certain demographic groups.

### 9.2 Complex and Evolving Trust Relationships

First, trust is context-dependent and varies by topic, time, or situation (e.g., someone may be trusted for movie recommendations but not for financial advice). Additionally, trust changes based on user behavior, interactions, and new information. Trust models must adapt to these changes without being outdated or biased by past interactions. Second, users interact across multiple online platforms. Therefore, building a cross-platform trust model and deriving an overall trust score is necessary, although difficult.

### 9.3 Privacy Concerns and Malicious Manipulation

Collecting and analyzing trust-related data may expose sensitive user information. Therefore, striking a balance between trust modeling and user privacy protection is crucial [129]. In addition, detecting fake accounts, bots [94, 123], and mitigating Sybil attacks [118] or reputation manipulation are ongoing challenges.

## 10 CONCLUSIONS

Trust modeling research in online social networks focuses on developing computational models and algorithms to simulate the formation, maintenance, or erosion of trust in online communities. Going further, it helps assess trust relationships, verify the credibility of information, detect malicious or spam activity, and foster more trustworthy and high-quality online communities. To define trust, trust is a willingness to accept vulnerability based on positive expectations of others. It stems from both rational evaluation and emotional bonds. Trust takes time to build, especially in anonymous online communities, but can be lost quickly. To simplify and facilitate trust modeling, we dissect the concept of trust and identify several factors that influence the formation and evolution of online trust, including competence, reciprocity, consistency, transparency, and similarity. Next, we systematically classify state-of-the-art trust quantification models into ten algorithmic categories. Various trust models collectively advance our understanding of how trust is established, maintained, or propagated. Subjective logic and fuzzy logic deal with the inherent uncertainty and ambiguity in trust relationships. Reputation-based trust models aggregate community opinions and accumulate trust scores based on behavior, contributions, and peer feedback during interactions. Graph-based trust models simulate social networks as weighted graphs, where users are nodes and trust relationships are edges. Trust is propagated or aggregated through connections and paths. Context-aware trust models are developed for specific domains, topics, or situations, rather than using a one-size-fits-all approach.

Research on trust models in online social networks is still in its emerging stage, and many issues remain unresolved. As part of our efforts, we provide an implementation-centric handbook on trust modeling and suggest potential research directions to encourage further exploration in this field. While building effective trust models is challenging, we believe that through in-depth research, trust can evolve from an ideal concept into a powerful tool in online interactions.